\documentstyle[12pt,aaspp4]{article}

\received{}
\accepted{}
\slugcomment{Submitted to Science}

\begin{document}
\include{psfig}
\def\kms{km~s$^{-1}$ }
\def\Lya{Ly$\alpha$ }
\def\lya{Ly$\alpha$ }
\def\Lyb{Ly$\beta$ }
\def\lyb{Ly$\beta$ }
\def\Lyg{Ly$\gamma$ }
\def\Lyd{Ly$\delta$ }
\def\Lye{Ly$\epsilon$ }
\def\d{$d_5$ }
\def\Ly{Lyman}
\def\ang{\AA }
\def\gq{$\geq$ }
\def\zem{$z_{em}$ }
\def\zabs{$z_{abs}$ }

% -- 1. Title page !
%
\title{Cosmological Deuterium Abundance and the
Baryon Density of the Universe} 

\author{SCOTT BURLES\altaffilmark{1} \& DAVID TYTLER\altaffilmark{1}} 
\affil{Department of Physics, and Center for Astrophysics and Space
Sciences \\
University of California, San
Diego \\
C0111, La Jolla, CA 92093-0111}
 
\altaffiltext{1}{Visiting Astronomer, W. M. Keck Telescope, California
Association for Research in Astronomy}
 
%
% -- 2.Abstract page!
%
 
\begin{abstract}
Standard big bang nucleosynthesis (BBNS) promises accurate predictions of
the primordial abundances of deuterium, helium-3, helium-4 and
lithium-7 as a 
function of a single parameter.  Previous measurements have nearly
always been interpreted as confirmation of the model
(\cite{cop95}).
Here we present a measurement of the deuterium
to hydrogen ratio (D/H) in a newly discovered high redshift metal-poor
gas cloud at redshift $z=2.504$. This confirms our earlier
measurement of D/H (\cite{tyt96}), and together they give
the first accurate measurement of the primordial D abundance,
and a ten-fold improvement in the accuracy of the
cosmological density of ordinary matter.
This is a high density, with most ordinary matter
unaccounted or dark, which is too high to agree with
measurements of the primordial abundances of helium-4 and lithium-7.
Since the D/H measurement is apparently simple, direct, accurate
and highly sensitive, we propose that helium requires a systematic correction,
and that population II stars have less than the primordial abundance of $^7$Li.
Alternatively, there is no concordance between the light element abundances,
and the simple model of the big bang must be incomplete and lacking physics,
or wrong.

\end{abstract}

%\section{INTRODUCTION}
 
In the standard big bang model the primordial abundances of
the light elements depend on the unknown value of the
cosmological baryon (ordinary matter)
to photon ratio, $\eta$ (\cite{cop95},\cite{wag67}--\cite{sar96}).
A measurement of primordial D/H, which is very sensitive to
$\eta$,
leads to precise predictions for the abundances of the other
light elements, which can be compared with observations to test the model.
 
Unfortunately D is destroyed inside stars, which then eject gas with H but no
D. As more gas is processed and ejected, the
value of D/H in the interstellar medium of our Galaxy drops further below
primordial. We look for primordial D/H
in quasar absorption systems (QAS) (\cite{ada76},\cite{web91}), 
some of which
sample gas at early epochs, and in intergalactic space where there
are too few stars to destroy significant amounts of D.
 
The absorption system at $z=2.504$ 
towards QSO 1009+2956 was 
from 40 QSO spectra which we obtained at Lick
Observatory because it showed a steep Lyman limit break
and weak metal lines, which indicate a low $b$-value 
suitable for the detection of deuterium.
Of the 15 QSOs which met these criteria and were 
subsequently observed with the 10-m W. M. Keck
telescope, only 2 have yielded deuterium measurements (\cite{this}).
We now describe a measurement of D/H, in this new QAS, which agrees
with our previous low value, but not with
measurements of a high D/H
(\cite{son94},\cite{car94}). Either
BBNS was inhomogeneous (\cite{jed95}) and D/H varies spatially, or
more likely, the spectra showing high D/H were contaminated.

In Figure 1 we show nine lines in the absorption system at $z=2.504$
towards QSO 1009+2956.
All six metal absorption lines are best fit by two components
separated by 11 km~s$^{-1}$, using the parameters given in Table 1.
The dotted lines show the expected positions of the H and D lines
for the gas at the well determined redshifts of these components.
The two components adequately fit the blue side of the Lyman features, but
an extra H component at $v_{rel}= +40$ \kms is seen on the red side
which is not significant in our measurement of D/H.
 
The D Lyman-$\alpha$ absorption is unsaturated
and has a well determined total column density, N(D~I).
However the amount of D in the red component is not well determined
because it is blended with both the blue D and H components.
If we do not include the red D component, systematic
under-absorption would occur at that velocity, so we
set (D/H)$_{red}$=(D/H)$_{blue}$=(D/H)$_{total}$.
 
All lines from the same gas will
have the same turbulent velocity dispersions $b_{tur}$,
but the thermal widths will depend on the mass of each ion:
$b_{therm} = 0.128\sqrt{Tm_p/m_{ion}}$ \kms, where $m_p$ is the proton mass.
We use the $b$-values of Si, C, and H in Table 1 to obtain the
temperatures and $b_{tur}$ of the two components. These values are
consistent with a single $b_{tur}$ for all lines, and allow us to
calculate that the red component of D should have
$b$(D~I) $=\sqrt{b_{therm}^2 + b_{tur}^2}$ = 15.4 \kms which we use as a
constraint on the fit because we cannot measure this value from the D line.
The widths of the H and D lines are dominated by thermal motions, so the lines
are accurately fit by Voigt profiles (\cite{tyt96},\cite{lev96}).
A simultaneous fit to the Lyman $\alpha$, $\beta$, and
$\gamma$ lines in the Keck spectrum and the Lyman continuum
absorption in the Lick spectrum (Figure 2)
gives D/H = $3.0 \, ^{+0.6}_{-0.5} \times 10^{-5}$ ($1\sigma$ random photon and
fitting errors).
 
A systematic increase in D/H comes from the chance superposition
of weak H absorption at the expected position of D.
To estimate this, we made noise-less model spectra with
D and H lines at the redshifts given by the metal lines.
We used the $b$ values and N(H~I) from Table 1, but we vary N(D~I).
For each N(D~I) we made $10^6$ spectra, and to each of these we added
random lines to simulate the \Lya forest, using the known distribution of
$z$, N(H~I) and $b$-values for forest lines (\cite{hu95}).
We then calculated the likelihood that the data came from the model
spectra for each D/H. 
Assuming that
all values of D/H are equally likely {\it a priori}, the expected or mean
D/H is Log (D/H)$_{exp} = -4.60 \, _{-0.04}^{+0.02}$, where the errors
are the standard deviations of the likelihood distribution above and 
below the mean.
The correction from the Monte Carlo simulations
changes our value from a formal upper limit to a measurement of D/H.
Additional systematic errors in D/H from fitting the continuum level are
estimated to be $\Delta$Log (D/H) = 0.06. We varied the continuum
by 2\% near \Lya (Figure 1), and by 10\% near the Lyman continuum
(Figure 2), and for each combination we re-fit the spectra.
Including the uncertainties due to systematics, we find 
\begin{equation}
Log \left({D \over H}\right)
= -4.60 \pm 0.08 \pm 0.06
\end{equation}
where the errors represent
the $1\sigma$ random error followed by the 
systematic errors from the continuum level. 
 
The blue side of the blue D \Lya line appears unblended,
and is fit with $b$(DI) = 15.7 $\pm$ 2.1 km s$^{-1}$. This
is consistent the value of $b = 13.5 \pm 0.5$ \kms predicted from the
$b_{therm}$ and $T$ given by the metal and H lines.
Less than 2\% of H \Lya lines have $b$-values this small or smaller
(\cite{hu95}), so it is
unlikely that this line is strongly contaminated with H.
Metal lines can be this narrow, although they are usually narrower with
$b < 10$ km s$^{-1}$. The D feature is
unlikely to be a metal line because metal lines of this
strength are nearly always accompanied by other lines, and the spectra
has been searched for these lines.
The {\it a posteriori} probability of a random H line with $b< 17.8$ and
$N > 12.6$ cm$^{-2}$, 
to account for $>0.5$ of the D line, and redshift within 20 \kms of that
of the metal lines is $< 8 \times 10^{-4}$.
A high correlation of ``satellite" components on these velocity
scales could increase the probability by as much as a factor of 3
(\cite{web91},\cite{hu95}).
 
The residual flux below the Lyman edge gives an accurate 
measure of
all H~I in this velocity region (Figure 2, \cite{ste90}).
There are no other \Lya lines within 5000 \kms of $z=2.50$ which
have H~I column densities $>10^{16}$ cm$^{-2}$, so that all of the
Lyman continuum absorption must be produced by gas in the $z=2.504$ absorption
system.
The blue side of the \Lya, \Lyb and \Lyg lines are best fit if all
of this H~I is near the two velocity components which are seen in the metals.
There could be be additional H at velocities between the metal lines and
$+40$ \kms, provided this gas has very low metal abundances ([C/H] $< -3.5$).
However, nearly all known QAS with large N(H~I) have metal
abundances [C/H] $> -3$.
 
The column densities of the metals and neutral hydrogen give
the metallicity and neutral fraction of the gas.
Following Donahue \& Shull (\cite{don91}),  we model the system as an optically
thin gas ionized by a typical QSO photoionizing
spectrum given by Mathews \& Ferland.
We estimate the ionization parameter (the ratio of the number of photons
with energies above one Rydberg to the number of atoms):
U $\equiv n_\gamma/n_p \approx 10^{-2.8}$,
which gives the metal abundances shown in Table 1.
For a photoionization model of low metallicity components,
this corresponds to a neutral hydrogen fraction
of H~I/H $\approx 10^{-2.5}$.
 
The measured D/H is consistent with normal Galactic chemical evolution
(\cite{tyt96},\cite{edm94});
the destruction of D in known populations of stars
can account for our D/H, that in the per-solar nebula, and the current
ISM D/H.
In the ISM, D/H $= 1.6 \pm 0.1 \times 10^{-5}$ (\cite{lin95}),
for [O/H] = -0.25. 
If the destruction of D is proportional to [O/H], then
we would expect 0.005 of the D would be destroyed for [O/H] $<-2.5$
(\cite{ste95}).
 
We find [C/Si] $\simeq -0.3$ in both components
which is characteristic of low metallicity stars in the halo of our Galaxy.
This suggests that the C and Si were created in ``normal" supernovae. If
some additional astrophysical processes destroys D, it must do so without
producing more C and Si than we see,
or other elements which we would have seen if they were made,
and without changing the usual C/Si ratio.
If primordial D/H were $24 \times 10^{-5}$ (\cite{son94},\cite{car94}) then
87\% of D must be destroyed in Q1009$+$2956 and 90\% in Q1937$-$8118.
The level of destruction would be large and similar for
[C/H] $= -3.0 $ to $ -2.2$, but small for Q0014+8118 with [C/H] $< -3.5$.
Redshift is apparently not a factor.
 
%\section{THE COSMOLOGICAL PRIMORDIAL D/H RATIO}

The QAS towards Q1937-1009 in which
we previouly measured a low D/H 
is extremely similar to Q1009+2956, except that
N(H~I) is larger, and the redshift is higher (\cite{tyt96}). We used
the Lyman series lines up to 19, and the
absence of flux in the Lyman continuum to constrain the N(H~I), and
we obtained smaller errors: Log D/H $= -4.64 \pm 0.06 \pm 0.06$.

The best estimate of the primordial D/H is obtained by taking the average of
our two measurements weighted by the squares of their random errors: 
\begin{equation}
Log \left({D \over H}\right) = -4.62 \pm 0.05 \pm 0.06,
\end{equation}
where the first error is the random error on the weighted mean,
and the second is the larger of the systematic errors from the
continuum level uncertainty.
We do not add these systematic errors because they have a similar origin
and the two measurments agree to within the random errors:
Prob$(\chi^2 \geq 0.18) = 0.67$.
In linear units, the mean is
\begin{equation}
\left({D \over H}\right) = 2.4 \pm 0.3 \pm 0.3 \times 10^{-5} 
\end{equation}

In Table 2 we list all published measurements of D/H in QAS.
Only the two discussed above are measurements.  The others were initially
presented as limits or possible detections, 
and they will
all be biased to higher than true D/H because they do not include
corrections for weak H at the position of D. 

The QAS towards Q0014+8118 differs from our two in several ways.
Since no metal lines were detected, 
the velocity structure of the cloud could be determined only by
median filtering of the higher-order lines in the \Lya forest (\cite{son94}).  
The Lyman-$\alpha$
feature was considerably more complex: five components were required
for an adequate fit (instead of three), 
with two components within 30 \kms of the deuterium
absorption line (\cite{car94}).  The neutral hydrogen column density of
the component where deuterium is measured, Log N(H~I) = 16.74, is
4 and 10 times lower than the column densities in our two 
QAS, which reduces the sensitivity to low D/H.

Rugers \& Hogan reanalyzed the published spectra of Q0014+8118,
and determined that the deuterium feature is better
fit with two very narrow components separated by 21 \kms (\cite{rug96}).
They claim that it is very unlikely that there are two narrow lines by chance at
the expected position of D, but we are not convinced 
because there are no metal lines to constrain the 
velocities and their model is a poor fit to 
the data in many places (\cite{poorfit}).
These
data remain consistent with this D line being contaminated (\cite{ste94}).

Towards 
QSO 1202-0725, Wampler et al. (\cite{wam96}) find $D/H < 15 \times 10^{-5}$
at a redshift $z=4.672$.  This QAS has high metallicity, [O/H] = 0.3,
and does not look suitable for inferences of primordial D/H.
Towards QSO 0420-3851, Carswell et al. (\cite{car96}) find a lower
limit of $D/H > 2 \times 10^{-5}$ at $z=3.086$.  This QAS also has
a high metallicity, [O/H] = -1.0.  The D~I column is fairly well-determined,
but the H~I is high, N(H~I) $> 10^{18}$ cm$^{-2}$, and is very uncertain.
All data in Table 2 are consistent with our low D/H value, but our data are
inconsistent with  high D/H, unless D/H is distributed inhomogeneously
(\cite{jed95}).

%\section{The Cosmological Baryon Density}

We now discuss the baryon density implied by our low D/H measurements.
In Figure 3 we show how D/H fits into the standard cosmological
framework.  For the remainder of this paper, we present the implications
of the new D/H measurements on these quantities. 
The current cosmological density of baryons, $\rho_b$, is given by
\begin{equation} 
\eta = 6.4 \, _{-0.4}^{+0.5} \, ^{+0.6}_{-0.5} \, ^{+0.3}_{-0.3} 
\times 10^{-10},
\end{equation} 
where the third error is the $1\sigma$ from nucleosynthesis
predictions (\cite{sar96}).
The density of photons from the Cosmic Microwave Background (\cite{mat94}),
$n_\gamma = 411 \, cm^{-3}$ gives:
\begin{equation}
\rho_b \equiv \eta \, n_\gamma \, m_p =  4.4 \, 
\pm 0.3 \, \pm 0.4 \, \pm 0.2 
\times 10^{-31} \, g \, cm^{-3},
\end{equation}
and as a fraction of the current critical density, $\rho_c = 3 H_0^2/8 \pi G$,
\begin{equation}
\Omega_b \equiv {{\rho_b} \over {\rho_c}}
= 0.024 \, \pm 0.002 \, \pm 0.002 \, \pm 0.001 \, h^{-2} 
\end{equation}
where the Hubble constant, $H_0 = 100 \, h$ \kms Mpc$^{-1}$.
Since the observed density of visible baryons in stars and hot gas is about
$\Omega_{LUM} = 0.003$ (\cite{per92}), 
most baryons (about 94\% for $h=0.7$) are
unaccounted.

%\section{OTHER LIGHT ELEMENTS}

In Figures 4--6, we present the predicted adundances
of $^4$He, $^3$He and $^7$Li relative to hydrogen 
(\cite{sar96}). We show values of
$\eta$ in the shaded region which are consistent with our D/H. 
The usually accepted estimates of primordial
$^4$He come from measurements
of low-metallicity extragalactic H~II regions.
Even modest chemical production can significantly change
the abundance of $^4$He, so the primordial value is 
inferred by extrapolating $^4$He measurements to zero metallicity
(\cite{pag92}--\cite{thu96}).
Pagel et al. reported an inferred primordial mass fraction of $^4$He,
$Y_p = 0.228 \pm 0.005$, with a 95\% upper bound
of 0.242 including systematic errors (\cite{pag92}).
Recently, Thuan et al. found $Y_p = 0.241 \pm 0.003$ (\cite{thu96}).

These $^4$He measurements are not consistent with our D/H, which predicts
\begin{equation}
Y_p = 0.249 \pm 0.001 \, \pm 0.001 \, \pm 0.001.
\end{equation}
To quantify the discordance of D/H with the $^4$He 
measurements mentioned above,
we form the weighted mean $\eta$ using our mean D/H and either of the
$^4$He values. We weight with the random errors alone, where we now
include nuclear uncertainties in quadrature in our D/H error.
We find Prob($\chi^2 > 5.2) = 0.02$ 
for the low $Y_p$, and 
Prob($\chi^2 > 3.0) = 0.08$ for the high $Y_p$.
If BBNS is correct, this disagreement shows that there must be systematic
errors which should be identified and corrected. 
Reducing our D/H by its systematic error, gives
Prob($\chi^2 > 3.7) = 0.05$ for the low $Y_p$ 
shifted up by its systematic error, 
and  Prob($\chi^2 > 2.7) = 0.1$ for the high $Y_p$, 
which does not have a quoted systematic error.
Even with the allowances of current systematic error estimates, 
D/H and Y$_p$ fail to predict the same value of $\eta$ with
BBNS. 

$^3$He is more complex and poorly understood
because it is both created and destroyed in stars.
The ratio of the D mass fraction in the ISM to primordial D
is $X_{D(ISM)}/X_{D0} = 0.67 \pm 0.09$ (random, not systematic error).
If $^3$He undergoes the same amount of destruction, then
its primordial abundance would be 1/0.67 times that in the ISM, which
is an upper limit because $^3$He is also created (e.g. 
primordial D is burned in stars to make $^3$He).
Low mass stars may produce $^3$He copiously
(\cite{roo76}), as is suggested by
its high abundance in the ejecta of planetary nebulae (\cite{roo92}).
If this is the case, only
upper limits on $^3$He, not D + $^3$He, can be
correctly applied to cosmology.  However the
observed abundance in Galactic H~II regions is 5 -- 20 times
lower than expected (\cite{gal95}), for unknown reasons,
perhaps because some of the $^3$He made in low mass stars is later
destroyed, or because
observations are made in H~II regions
which contain the ejecta of high mass stars which destroy
$^3$He (\cite{oli95b}).

The abundance of $^7$Li in stars in the disk of our Galaxy
(population I) spans three orders of magnitude, because Li is both
made (e.g. by cosmic ray spallation in the interstellar medium, and
in novea and supernovae) and destroyed, but
warm metal-poor halo (population II) stars have similar
``plateau" abundances of $^7$Li in their atmospheres.
These measurements (\cite{spi82}-\cite{tho94}), shown in Figure 6,
are often taken as the primordial abundance, but
recent data show that this is not justified.
Stars with very similar temperatures and metallicities,
subgiant stars in the globular cluster M92,
and turnoff stars in the halo, have significantly different
$^7Li$ abundances (\cite{rya96}), perhaps because of
different amounts of $^7$Li depletion.  Ryan et al. have found that
the $^7$Li abundance depends on both temperature and metallicity in
ways which were not predicted and are not understood.
If we are to obtain a primordial $^7$Li abundance we must
either (1) understand why its abundance varies from star to star, and learn to
make quantitative predictions of the level of depletion, or (2) make
measurements in relatively unprocessed gas.
Our D/H measurements imply that $^7$Li in population II stars has been
depleted by about 0.5 dex, which may be difficult to reconcile with the
near constancy of $^7$Li in the warm halo stars.

We conclude that our D/H measurements are probably the first measurements
of a primordial abundance ratio of any elements, and that they
give about an order of magnitude improvement in the accuracy of
estimates of $\rho_b$.

Where are the baryons?
The MACHO collaboration recently announced detection of baryonic
dark matter in the halo of our Galaxy from
7 microlensing events towards the Large Magellanic Cloud (\cite{pra96}). 
The most likely mass of the MACHOs is 0.3 - 0.5 solar masses, from 
the durations of the events and the velocities implied by the model of 
the halo of our Galaxy. This corresponds to a standard halo model mass of 
1.6$\times 10^{11}\,M_\odot$, and a large fraction of the total mass.
If the halos in all galaxies are entirely baryonic MACHOS,
they have $\Omega_{MACHO} > 0.13$ (90\% confidence, with prefered values
exceeding 0.3 for h=0.75 \cite{zar94}). This is more than 
$\Omega_b = 0.043 \, \pm 0.009$, which is predicted from our two measurements,
so $<0.5$ (prefered value 0.16) of halos are baryonic.
The MACHO detections strongly imply a high $\Omega_b$: if halos are 50\% 
baryonic MACHOS then $\Omega_b > 0.065$ (90\% confidence for h=0.75),
enough to account for all baryons.

X-ray observations reveal that 
baryons in the form of hot gas contribute a fraction
$f_x = (0.05 - 0.14) \, h^{-3/2}$ of the 
total mass of clusters of galaxies (\cite{whi93}).
The remaining mass is some combination of unseen dark baryons (e.g. MACHOS)
and any other non-baryonic dark matter, such as massive supersymmetric 
particles. The baryon fraction in clusters is then $f_b \geq f_x$.
Since galaxy clusters are the largest bound structures in the universe,
their $f_b$ should provide good estimates of
the cosmological value, so the
total mass density of the universe is then:
\begin{equation}
\Omega_{total} = \Omega_b /f_b \leq (0.14 -- 0.58) \, h^{-1/2},
\end{equation}
where we used $\Omega_b$ from D/H and the inequality allows 
for $f_b \geq f_x$ (\cite{afootnote}).
Although our D/H measurements give a very high $\Omega_b$,
it is not high enough to permit $\Omega_{total}=1$ from matter:
we have $\Omega_{total} < 0.72$ from $\Omega_b =0.029 \, h^{-2}$,
$h=0.6$, and $f_b = f_x$.
Either the universe is open, as is now allowed by
the theory of inflationary cosmology
$\Omega_{total} \leq 1$ (\cite{lin96}), or there is a cosmological
constant.

%\section{Conclusions}

We are forced to conclude, given the present state of primordial
abundance measurements, that D/H gives the only reliable constraints
on $\eta$ and $\Omega_b$.  The practice of finding consensus among
all the light elements should be heavily weighted to D/H, 
because it is most sensitive to $\eta$, and because of the apparent
absence of systematic uncertainties arising from destruction
and creation.  The discordance with other light elements
does not demand alternative models to the standard big-bang, 
rather it begs for all measurements of primordial abundances
to be made in similar pristine sites.  The search for more
deuterium measurements in QAS will continue, and at the current rate,
$\Omega_b\,h^2$ is likely to be known to 5\% by the end of the millenium.

\clearpage

%****************** STANDARD TABLE *******************************
% for centering Table 1

\begin{table*}   
\begin{center}
\begin{tabular}{cccc}
Ion & Blue Component & Red Component & Total \cr
\tableline
H I & N = 17.36 $\pm$ 0.09 & N = 16.78 $\pm$ 0.11 & N = 17.46 $\pm$ 0.05 \cr
& b = 18.8 $\pm$ 0.5 & b = 21.9 $\pm$ 4.1 & \cr
D I & N = 12.84 $\pm$ 0.09 & N = 12.26\tablenotemark{*}
& N = 12.94 $\pm$ 0.06 \cr
& b = 15.7 $\pm$ 2.1 & b = 15.4\tablenotemark{\#} & \cr
Si III & N = 12.81 $\pm$ 0.06 & N = 12.55 $\pm$ 0.04 & N = 13.00 $\pm$ 0.03 \cr
    & b = 4.9 $\pm$ 0.4 & b = 4.8 $\pm$ 0.8 & \cr
Si IV & N = 12.50 $\pm$ 0.03 & N = 12.05 $\pm$ 0.08 & N = 12.63 $\pm$ 0.02 \cr
    & b = 4.9 $\pm$ 0.6 & b = 3.9 $\pm$ 1.5 & \cr
C II & N = 12.46 $\pm$ 0.12 & N = 12.18 $\pm$ 0.18 & N = 12.64 $\pm$ 0.08 \cr
    & b = 7.0 $\pm$ 3.2 & b = 3.7 $\pm$ 4.2 & \cr
C IV & N = 12.81 $\pm$ 0.04 & N = 12.56 $\pm$ 0.06 & N = 13.00 $\pm$ 0.03 \cr
    & b = 5.4 $\pm$ 0.6 & b = 5.6 $\pm$ 1.2 & \cr
T($10^4$ K) & 2.1 $\pm$ 0.1 & 2.4 $\pm$ 0.7 & \cr
b$_{tur}$ & 3.2 $\pm$ 0.4 & 2.3 $\pm$ 1.4 & \cr
[C/H]     & $-2.9$ & $-2.8$ & ... \cr
[Si/H]     & $-2.5$ & $-2.6$ & ... \cr
[C/Si]     & $-0.4$ & $-0.2$ & ... \cr
 
\tablenotetext{*}{We fixed this column density 
to make D/H 
the same value in both components.}
\tablenotetext{\#}{This $b$-value is calculated from the most likely turbulent
velocity and temperature of the red component.}
 
\end{tabular}
\end{center}
 
\tablenum{1}
\caption{Model parameters for QAS at $z=2.504$ towards Q1009+2956.
Column densities (N) are in logarthmic units of cm$^{-2}$, while the
$b$-values are in units of \kms.}
 
\end{table*}
 
\clearpage
 
%****************** STANDARD TABLE *******************************
% for centering
% 
\begin{table*}   
\begin{center}
\begin{tabular}{ccccccc}
QSO & $z_{abs}$ & D/H$\times 10^5$ & 1$\sigma$ \tablenotemark{*} &
logN(H~I) & [C/H]\tablenotemark{\#} & Reference \cr
\tableline
ISM     & 0.0   & 1.6 & 0.1 & 18.2 & ... & \cite{lin95} \cr
1009+2956 & 2.504 & 2.5 & 0.5 & 17.46 & --2.9 & This Paper \cr
1937$-$1009 & 3.572 & 2.3 & 0.3 & 17.94 & $-$2.2, $-$3.0 & \cite{tyt96} \cr
0014+8118 & 3.320 & $\leq 19-25$ & ... & 16.7 & $<-$3.5 & \cite{son94},\cite{car94},\cite{rug96} \cr
1202$-$0725 & 4.672 & $\leq 15$ & ... & 16.7 & ... & \cite{wam96} \cr
0420$-$3851 & 3.086 & $\geq 2$ & ... & $\geq 18$ & -1.0 & \cite{car96} \cr  
\tablenotetext{*}{approximate standard deviation on D/H measurement}
\tablenotetext{\#}{Carbon to hydrogen ratio in logarthimic units, relative to solar}
\end{tabular}
\end{center}
 
\tablenum{2}
\caption{Recent Measurements of D/H}
 
\end{table*}

\clearpage

\section{Figure Captions}

Figure 1: Velocity plots of Lyman $\alpha, \beta, \gamma$ (left), and
all the metal lines
(right) in the absorption system towards QSO 1009+2956 ($z_{em}=2.616$,
V=16).
Zero velocity corresponds to the redshift $z = 2.503571$, of the
blue component.
The vertical dotted lines also show the red component at 2.503704.
The histogram represents the observed counts in each pixel normalized to the
quasar continuum.  The smooth curve show the best fitting Voigt profiles
convolved with the instrumental resolution (\cite{voigt}).
The Lyman lines and
Si~III (1206) are in the Lyman alpha forest region where there is
additional absorption which we do not fit.
 
On the nights of December 27, 28 1995, we obtained
5.9 hours of spectra with the HIRES echelle spectrograph on the
10-m W. M. Keck Telescope (\cite{vog94}).
Two exposures of 2.5 and 2 hours covered
3540 -- 5530 \AA, while a third 1.4 hour covered 3165 -- 4370 \AA.
A 1.14 arcsec slit produced spectra with resolution of
8 kms$^{-1}$.  Each exposure was accompanied by
dark, quartz lamp, and Throium-Argon arclamp exposures.  A standard star
was observed to trace the echelle
orders and remove the blaze response in the spectra.
We used our standard data reduction (\cite{tyt96}).
 
Figure 2: The Lick Spectrum of QSO 1009+2956 shows the Lyman Limit
due to the absorber at $z=2.504$.
On November 28, 1995 we used the Kast spectrograph on the Lick 3-m telescope
to obtain 1.9 hours of integration covering 3100 \AA - 5950 \AA.
The spectra were calibrated to vacuum heliocentric wavelengths, and
optimally extracted.
 
The smooth line shows the convolved
Voigt proflies of the higher order Lyman lines
calculated with the parameters in Table 1.  The optical depth, measured by
the ratio of flux blueward and redward
of 3200 \AA, constrains the total column density of neutral hydrogen.
 
Figure 3: A representation of the flow of information
in the standard cosmological model.
Boxes show measurements, and circles show theories and derived
quantities.  Most references to quantities shown are discussed
in the text.  The range for the Hubble Constant, $H_0$, is taken
from Mould et al. (\cite{mou95}).
When two numbers are shown, one is for $h=0.98$ and the other is
$h=0.64$. When $\Omega_b$ is used, we add or subtract all three errors
to enlarge the range.
 
Figure 4: The BBNS predicted primordial abundances of D and $^3$He
as a function of $\eta$ and $\Omega_b\,h^2$
(\cite{sar96}). 
The lower two rectangles are defined by the $1\sigma$ random plus
systematic errors on D/H towards Q1009+2956 and Q1937--1009.
The shaded region is from our mean D/H value, with $1\sigma$ random errors
from the quadratic sum of our random error plus the $\sigma$ nuclear
error. We then add on the systematic error.
The upper limit shows D/H towards Q0014+8118.

Figure 5: As Figure 4 but for 
$Y_p$, the mass fraction of $^4$He.
Dashed rectangles show the bounds (1$\sigma$ statistical plus systematic
errors) from recent measurements
of $^4$He (\cite{pag92},\cite{thu96}). The lack of 
intersection with the shaded region illustrates the inconsistency of
D/H with Helium-4 measurements.  The lower rectangle represents the
estimates of $^4$He deduced by Pagel et als., and it includes in its range
a systematic underestimation of $Y_p$ of 0.004.

Figure 6: As Figure 4 but $^7$Li/H 
The Spite and Thorburn $^7$Li plateau from measurements of 
population II stars are shown as dashed and
dot-dashed lines respectively.
(\cite{spi82},\cite{tho94}).  Significant depletion of
the surface $^7$Li abundance is the likely source of the discordances.
(\cite{rya96}).  The dotted line shows upper limits from lithium measurements
in population I stars.

\begin{figure}
\figurenum{1}
\centerline{
\psfig{figure=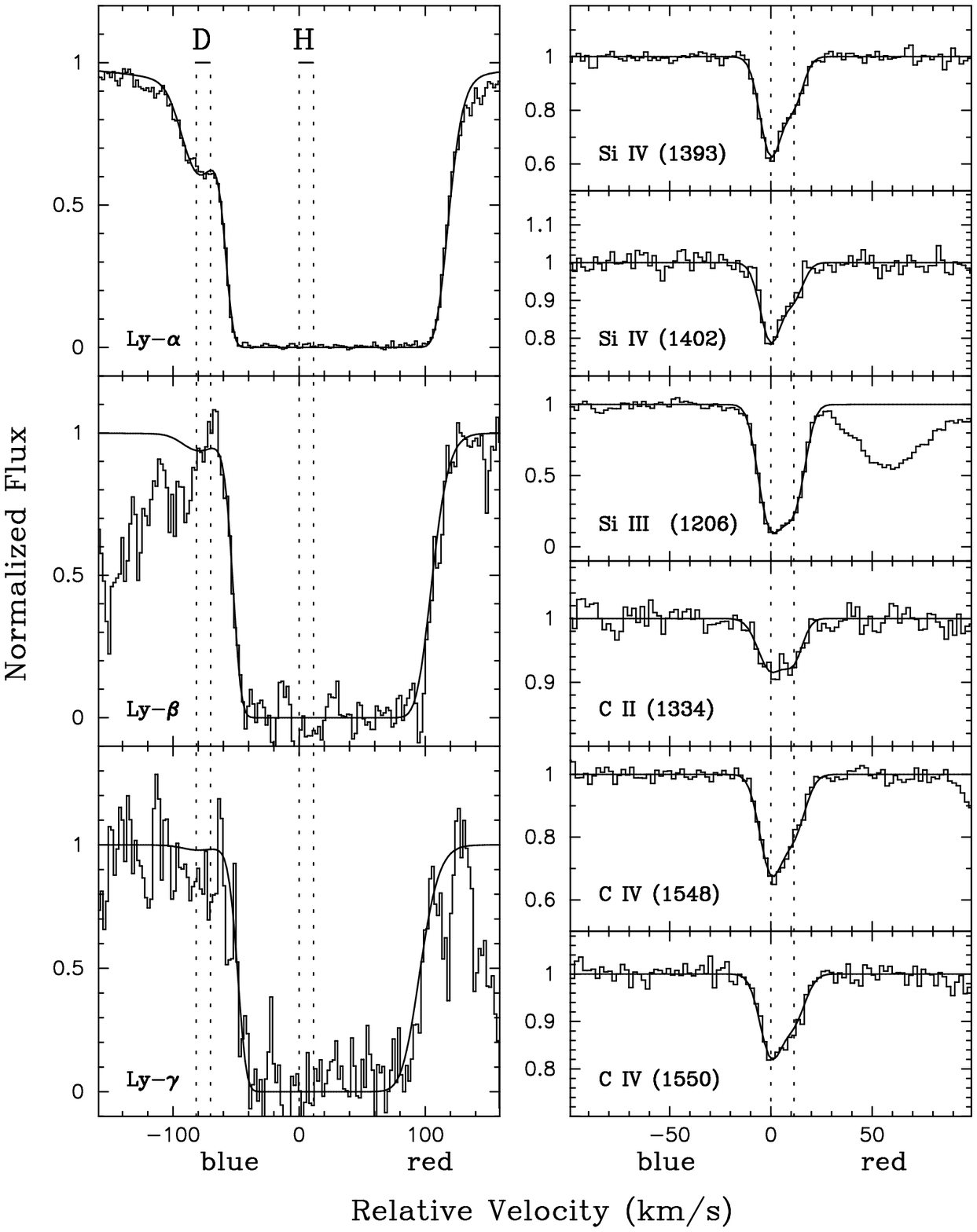,height=9.5in}}
\end{figure}
 
\begin{figure}
\figurenum{2}
\centerline{
\psfig{figure=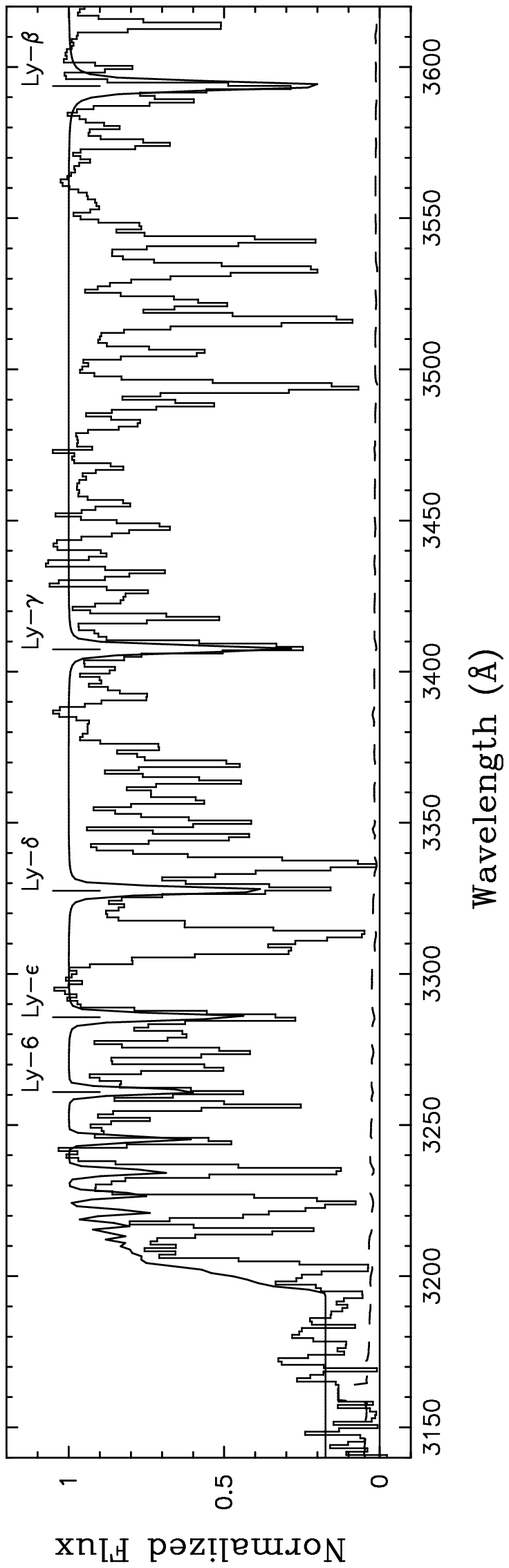,height=9.5in}}
\end{figure}

\begin{figure}
\figurenum{3}
\centerline{
\psfig{figure=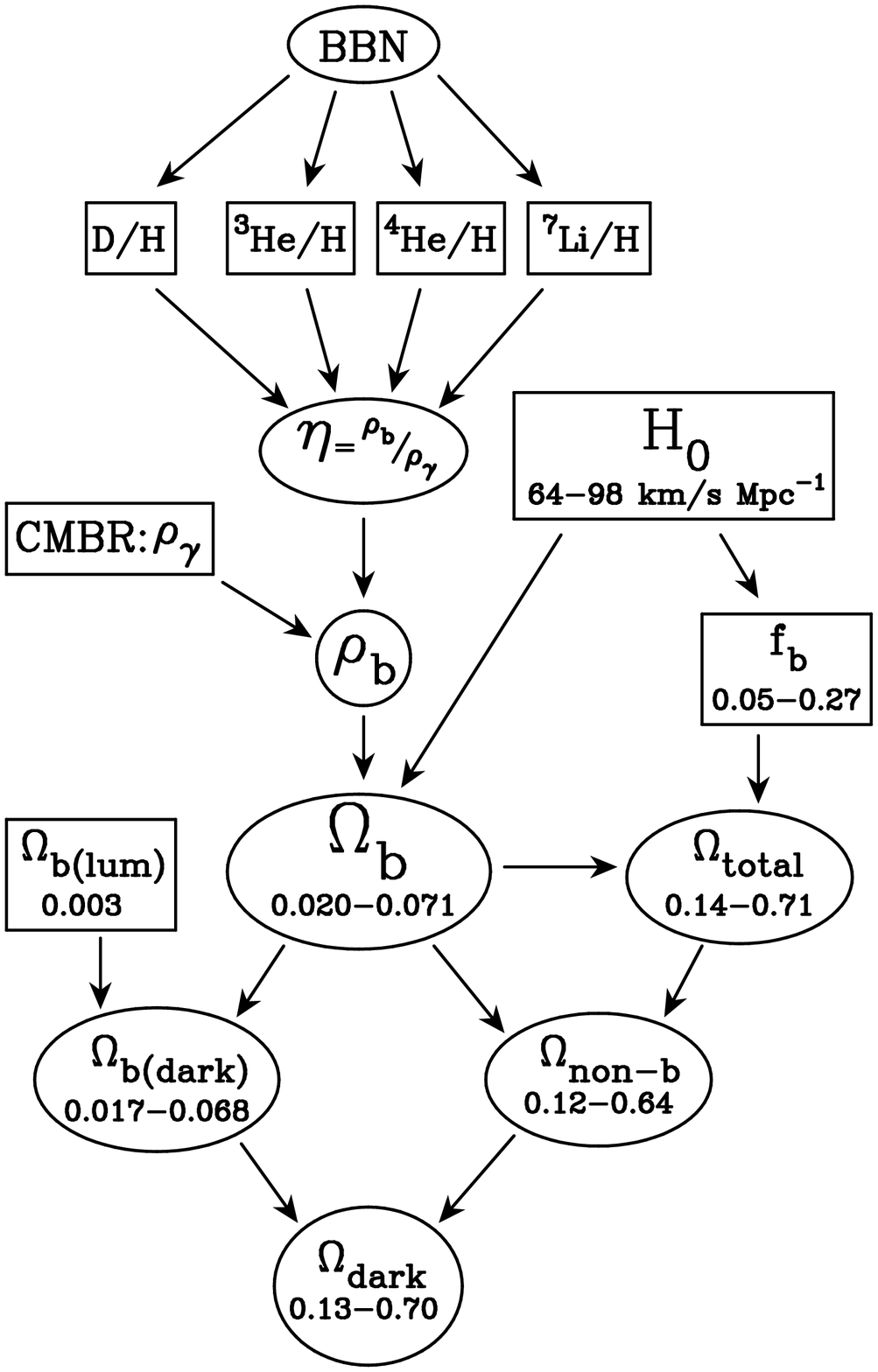,height=9.0in}}
\end{figure}

\begin{figure}
\figurenum{4}
\centerline{
\psfig{figure=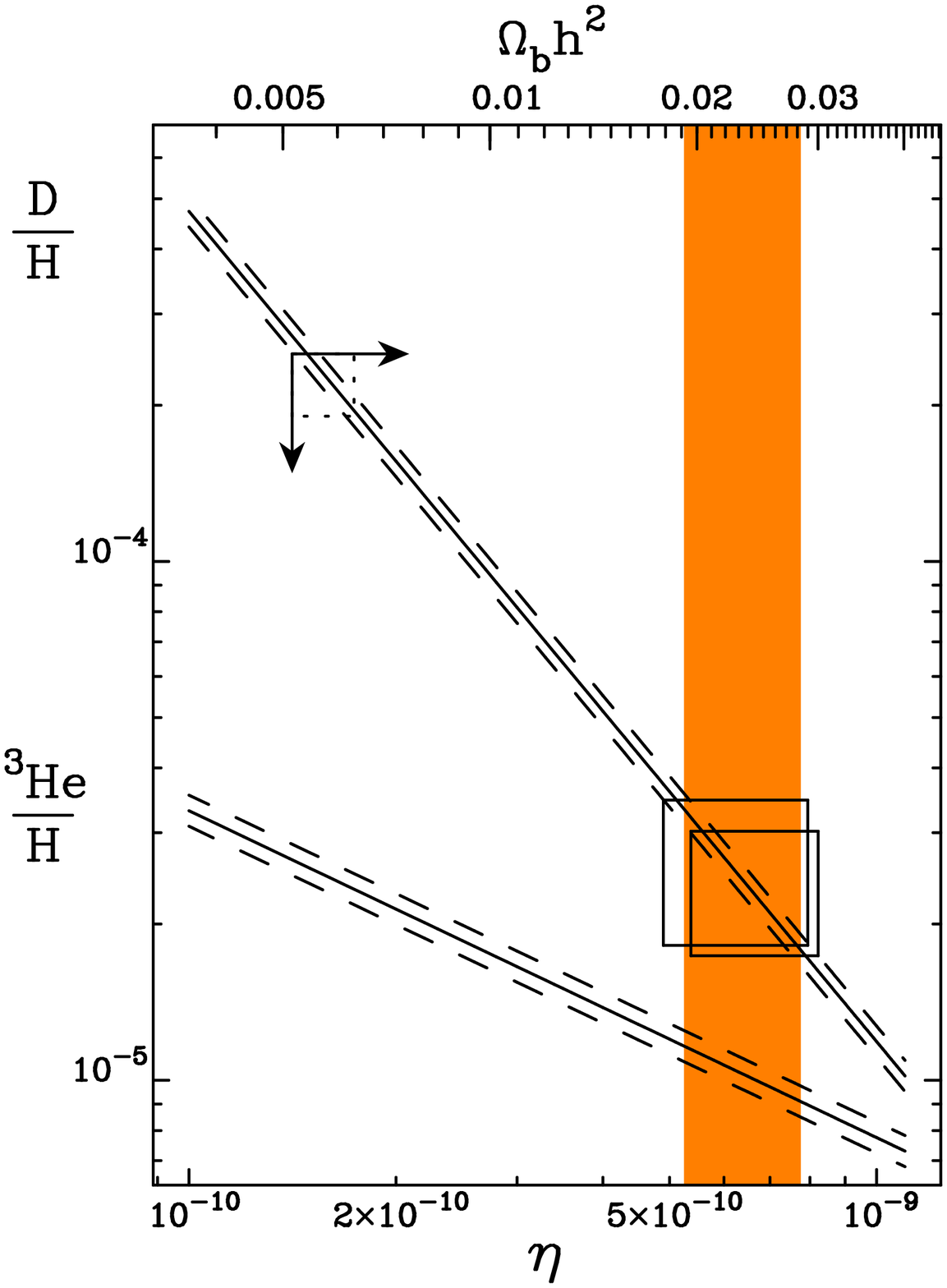,height=9.0in}}
\end{figure}
 
\begin{figure}
\figurenum{5}
\centerline{
\psfig{figure=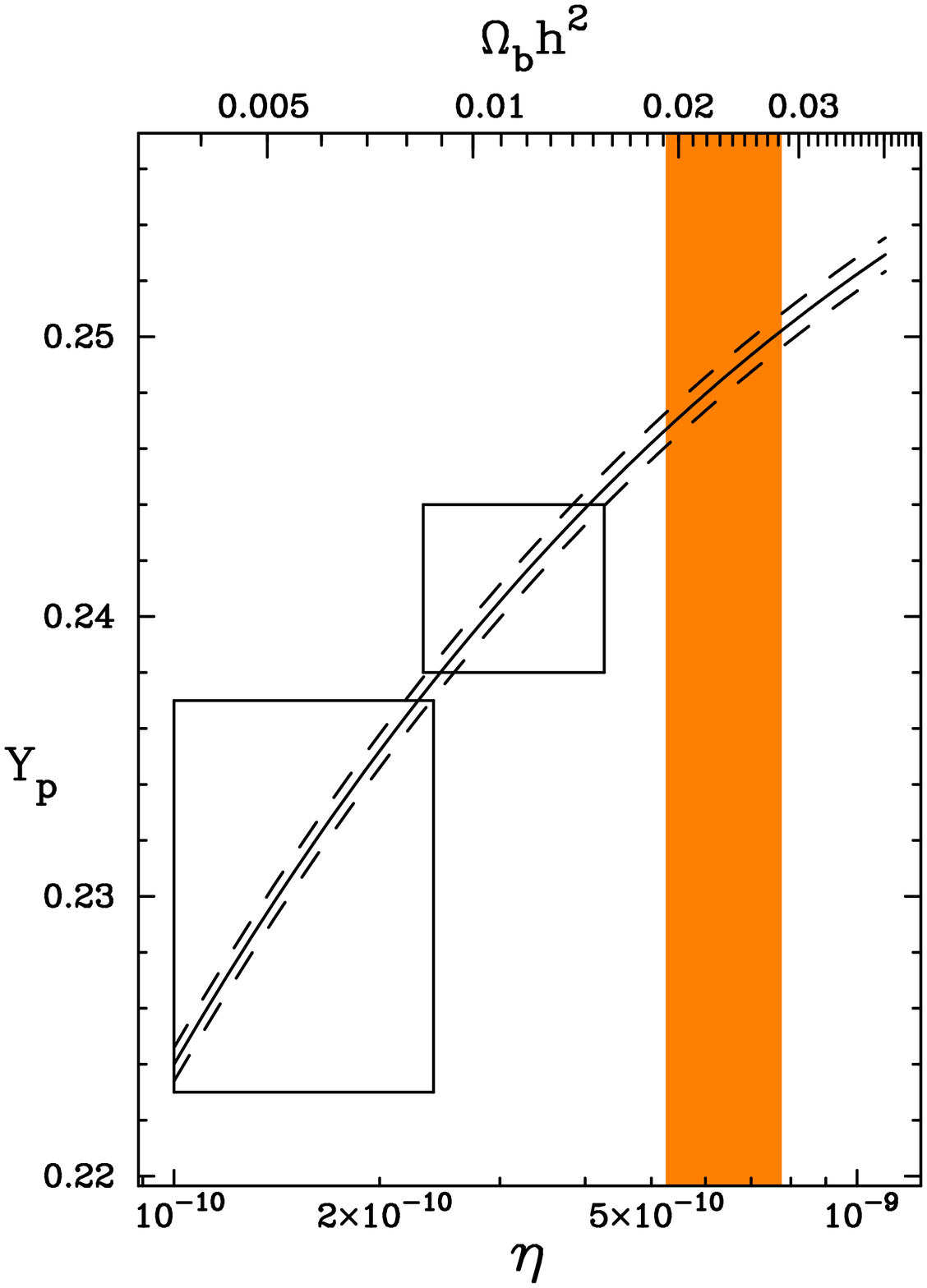,height=9.0in}}
\end{figure}
 
\begin{figure}
\figurenum{6}
\centerline{
\psfig{figure=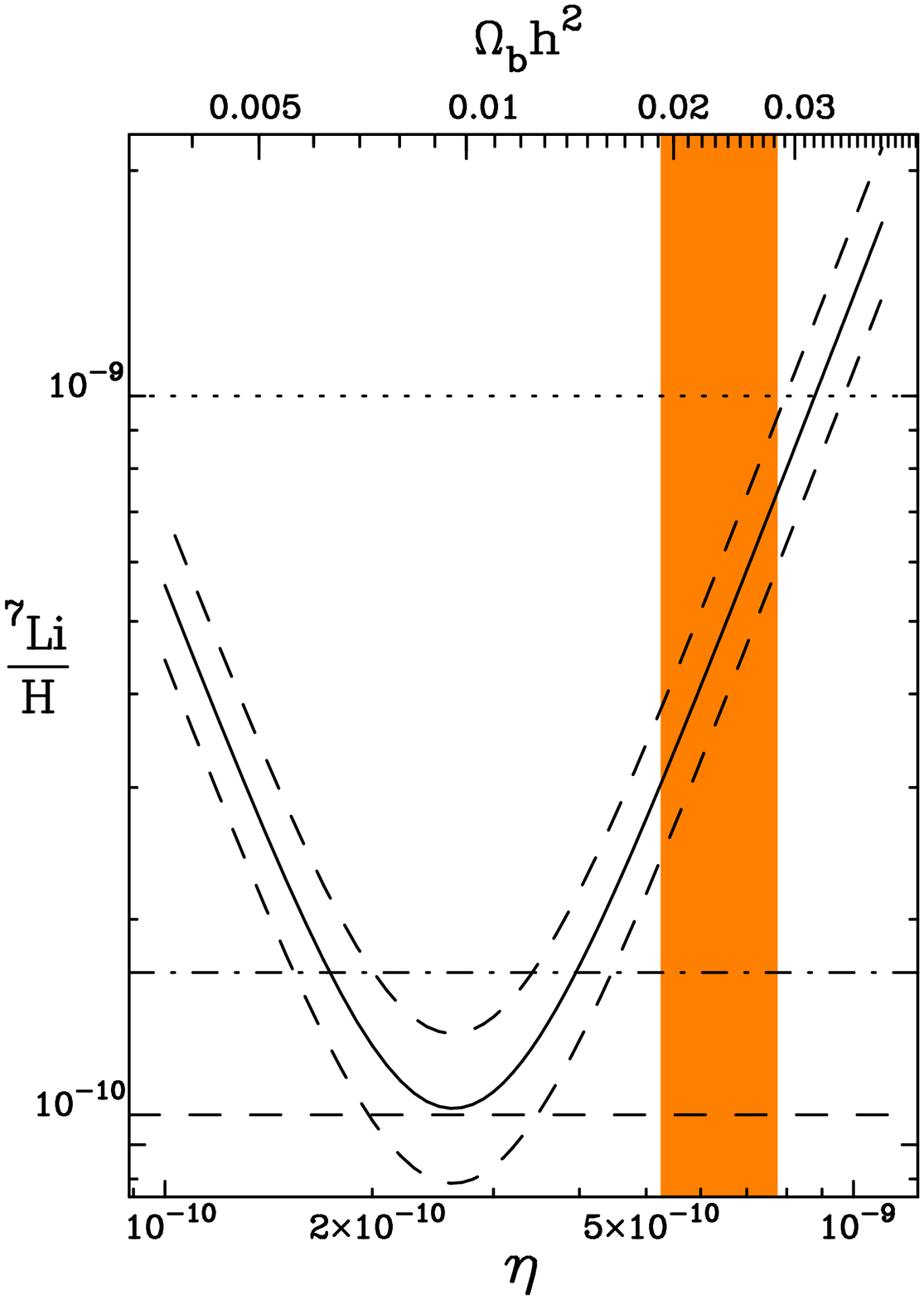,height=9.0in}}
\end{figure}
 

\begin{thebibliography}{}
\bibitem[8]{this}
8. Our first D/H measurement is reported in (2), our second measurement
is the subject of this article.
\bibitem[20]{poorfit}
20. The model presented by Rugers \& Hogan (\cite{rug96}) is a poor
fit to the data in several places: D~I \Lya, H~I \Lyg and \Lyd blue
wing and central spike, and high order lines at 3974, 3972, 3962.5, 
3961,3959.5, 3954.8 \AA.
\bibitem[40]{afootnote}
40. We add or subtract all three errors to give the
largest range for $\Omega_b$.
\end{thebibliography}
\end{document}